\documentclass[aps,twocolumn,showpacs]{revtex4}
\usepackage{psfig}

\newcommand{\be}{\begin{equation}}
\newcommand{\ee}{\end{equation}}
\newcommand{\ba}{\begin{eqnarray}}
\newcommand{\ea}{\end{eqnarray}}

\begin{document}

\title{
%
%preprint number:
%
\[ \vspace{-2cm} \]
\noindent\hfill\hbox{\rm  } \vskip 1pt
\noindent\hfill\hbox{\rm Alberta-Thy 16-01} \vskip 1pt
\noindent\hfill\hbox{\rm SLAC-PUB-9084} \vskip 1pt
\noindent\hfill\hbox{\rm hep-ph/0112117} \vskip 10pt
%
% now the title:
%
%\title{
Pion pole contribution to hadronic light-by-light scattering and 
muon anomalous magnetic moment
}

\author{Ian Blokland and Andrzej Czarnecki}
\affiliation{
Department of Physics, University of Alberta\\
Edmonton, AB\ \  T6G 2J1, Canada\\
E-mail: blokland@phys.ualberta.ca,  czar@phys.ualberta.ca}

\author{Kirill Melnikov}
\affiliation{Stanford Linear Accelerator Center\\
Stanford University, Stanford, CA 94309\\
E-mail: melnikov@slac.stanford.edu}

\begin{abstract}
We derive an analytic result for the pion pole contribution to the
light-by-light scattering correction to the anomalous magnetic moment
of the muon, $a_\mu = (g_\mu-2)/2$. Using the vector meson dominance model
(VMD) for the pion transition form factor, we obtain $a_\mu^{{\rm
LBL},\pi^0} = +56 \times 10^{-11}$.
\end{abstract}

\pacs{13.40.Em,12.40.Vv}
\maketitle

The recent measurement of the muon anomalous magnetic moment of the muon 
by the E821 experiment in Brookhaven \cite{Brown:2001mg},
\be
a_{\mu} = 116~592~020(160) \times 10^{-11}
\label{expres}
\ee
significantly deviates from the Standard Model.
Theoretical predictions, which include five-loop QED and two-loop
electroweak effects, rely on experimental data and theoretical models
to account for the non-perturbative hadronic contributions.  Depending
on the treatment of the latter, the discrepancy between the experiment
and the theory can be as large as $2.6\, \sigma$.  After the release
of the E821 result (\ref{expres}), the hadronic effects came under
renewed scrutiny and the reliability of various estimates has been
disputed \cite{Marciano:2001qq,Melnikov:2001uw}.  The main focus of
those discussions was the hadronic vacuum polarization which modifies
the photon propagator and has been evaluated using data on $e^+e^-$
annihilation into hadrons and the $\tau$ lepton hadronic decays
\cite{Eidelman:2001ju,Jegerlehner:2001wq,Hocker:2001fu}.
 
\begin{figure}[htb]
\hspace*{-38mm}
\begin{minipage}{16.cm}
\begin{tabular}{cc}
\psfig{figure=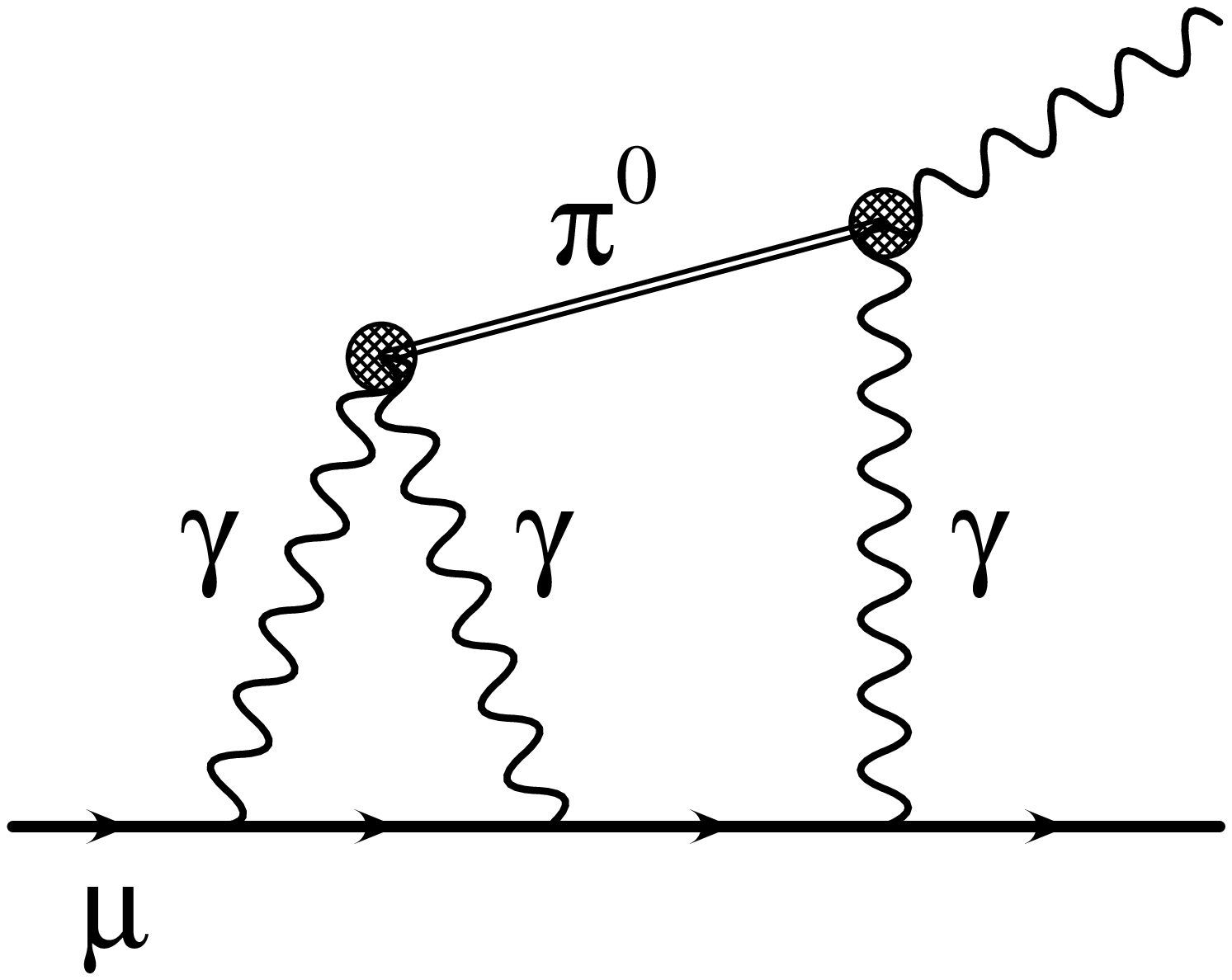,width=40mm}
&\hspace*{0mm}
\psfig{figure=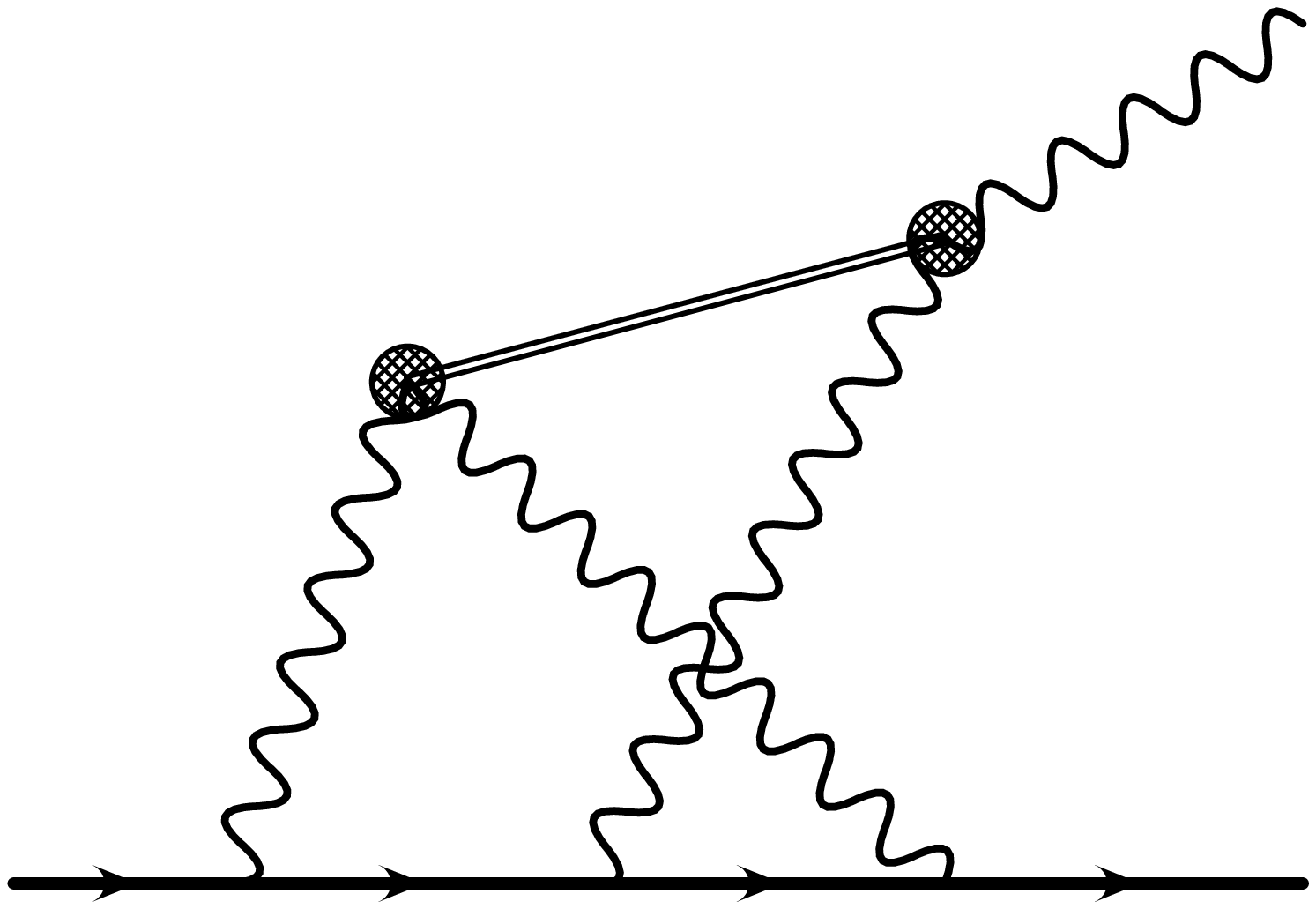,width=40mm}
\\
(a) & (b)
\end{tabular}
\end{minipage}
\caption{Pion pole contributions to the muon $g-2$.}
\label{fig:giraffe}
\end{figure}

Very recently it has been pointed out
\cite{Knecht:2001qf,Knecht:2001qg} that a significant part of the
discrepancy between the theory and the experiment may be due to the
theoretical evaluation of the pion pole contribution to hadronic
light-by-light scattering, which influences $g_\mu-2$ via diagrams
shown in Fig.~\ref{fig:giraffe}.  Namely, it has been claimed that
although the magnitude of those contributions computed in
\cite{Hayakawa:1995ps,Hayakawa:1996ki,Hayakawa:1998rq,%
Bijnens:1995cc,Bijnens:1996xf} is correct, they had been taken with an
incorrect (negative) sign.  

Since the pion pole contribution is the largest among the hadronic
light-by-light scattering contributions to $a_\mu$, the change in sign
has important implications for the comparison of the experimental result
with the theoretical prediction.  If the correct sign is positive, the
theory and experiment are in much better agreement (about
$1\,\sigma$ is removed from the reported discrepancy).  Motivated by
this, we have recalculated the pion pole contribution to the
light-by-light scattering correction to $a_\mu$.  Using the vector
dominance model for the pion transition form factor, we obtain
\be
a_\mu^{{\rm LBL},\pi^0} = +56 \times 10^{-11}, 
\label{res}
\ee
thereby confirming the result of Ref.~\cite{Knecht:2001qf}.
The purpose of this Letter is to describe the calculation leading
to Eq.~(\ref{res}).  We do not address the validity of
the model.  Our objective is to obtain an analytical (rather
than numerical) result and settle the issue of its sign.

At low energies,  the interaction of a neutral pion with photons is  described 
by the Wess-Zumino-Witten Lagrangian,
\be
{\cal L}_{\rm WZW} = -\frac {\alpha N_c}{12 \pi F_\pi} 
F_{\mu \nu} \widetilde F^{\mu \nu} \pi^0, 
\label{wzw}
\ee
where $N_c=3$ is the number of colors and $F_\pi\simeq 92.4$ MeV is the
pion decay constant.  Since this is a non-renormalizable interaction,
employing it in loop calculations
results in ultraviolet divergences.  While this is not a problem in
principle, since  divergent contributions can be absorbed into higher
dimensional operators of the chiral Lagrangian, in practice this precludes
any numerical estimate of the corresponding contributions because, as 
in the case of  muon anomalous magnetic moment, the relevant
counterterms are not known. In this situation one resorts to models to
obtain a finite result.  A simple and commonly adopted option
is to introduce a form factor into the $\pi^0 \gamma \gamma$ interaction
vertex, which damps the contributions of
highly virtual photons. This results in the following $\pi^0
\gamma \gamma$ interaction vertex:
\be
\frac {\alpha N_c}{3 \pi F_\pi} F_{\pi^0 \gamma \gamma}(q_{1}^2,q_{2}^2) 
i\epsilon_{\mu \nu \alpha \beta} q_1^{\alpha} q_2^{\beta},
\ee
where $q_{1,2}$ denote the momenta of the two outgoing photons.

The transition 
form factor $F_{\pi^0 \gamma \gamma}(q_{1}^2,q_{2}^2)$ depends on the
adopted model.  For our purpose it is sufficient to use the simplest
of these models (VMD) where it is assumed that the transition form factor 
reads
\be
F_{\pi^0 \gamma \gamma}(q_{1}^2,q_{2}^2) = \frac{M^2}{M^2 - q_1^2} 
\frac{M^2}{M^2 - q_2^2},
\label{formf}
\ee
where the parameter $M$ is phenomenologically determined to be 
close to the mass of the $\rho$ meson $M \approx m_\rho \simeq 769$ MeV.

The simple structure of the form factor opens a way for analytic
calculations. 
Initially, the diagram has three widely separated scales:
\ba
m_\pi^2 - m_\mu^2 \ll m_\mu^2 \ll M^2,
\ea
and one can take advantage of this hierarchy to construct an expansion in
$\delta\equiv (m_\pi^2 - m_\mu^2)/m_\mu^2$ and  
$m_\mu^2/M^2$ using asymptotic expansions. 
The expansion in $\delta$ is especially simple and reduces to 
the Taylor expansion of the pion propagator. The expansion in $m_\mu^2/M^2$
is the so-called Large Mass Expansion
\cite{Chetyrkin91,Tkachev:1994gz,Smirnov:1995tg}. 

Consider first the diagram in Fig.~\ref{fig:giraffe}(a).  The lines
denoted by $\gamma$-propagators are in fact products of the massless
photon propagators $1/k_i^2$ and of the form factor terms
$M^2/(k_i^2-M^2)$.  The vector meson mass $M$ in the latter sets
the hard scale for the momentum integrals, whereas the muon mass sets
the soft scale.  (After the Taylor expansion in $\delta$
these are the only mass scales present.)  We can now obtain four
possible combinations of soft/hard integration momenta in the two
loops of Fig.~\ref{fig:giraffe}(a).  The leading quadratic logarithm
of the muon and $\rho$ mass ratio is obtained by adding contributions
with both momenta soft, both momenta hard, and the one with the
momentum in the triangle (left-hand side) loop hard, and the
right-hand side loop soft.  The fourth, soft-hard combination
contributes only in the next order, suppressed by $m_\mu^2/M^2$. 

The non-planar diagram in Fig.~\ref{fig:giraffe}(b) can be evaluated
in a similar manner, with the only difference that a fifth momentum
region appears.  Namely, the two integration momenta may be hard, but
their difference in the pion propagator may be soft.  In this case the
diagram factorizes into a product of two one-loop vacuum diagrams.

Integrations in all four regions can be carried out analytically, and
coefficients in both expansion parameters,
$m_\mu^2/M^2$ and $\delta$, can be obtained to an
arbitrary order.  Technically, the calculations involve integrations
of two-loop single-scale diagrams of the type of either a muon
propagator or a massive vacuum diagram.  The most complicated topology
arises when both loop momenta are soft, of order $m_\mu$. This leads
to two basic on-shell two loop diagrams of the self-energy type which 
can be evaluated using the recent results of Ref.~\cite{Fleischer:1999hp}.

The result can be written as
\be
a_{\mu}^{{\rm LBL},\pi^0} = \left ( \frac{\alpha}{\pi} \right )^3 
\frac{m_\mu^2}{F_\pi^2} \left ( \frac{N_c}{\pi} \right )^2  X_{\pi^0},
\label{respi0}
\ee
and we find 
\ba
\lefteqn{ X_{\pi^0} = \frac{1}{48}L^2
     + \left ( \frac{1}{96}  - \frac{\pi}{48\sqrt{3}} \right ) L
       - \frac{277}{10368} }
\nonumber \\
&&\quad + \frac{\pi}{24 \sqrt{3}}S_2 
 - \frac{17 \pi}{3456 \sqrt{3}}
          + \frac{19}{128}S_2  - \frac{\zeta_3}{288}
        - \frac{11 \pi^2}{15552} 
\nonumber \\
&&
       + \frac{m_\mu^2}{M^2}  
\left[
\frac{155}{1296} L^2
 - \left ( \frac{65}{1296} + \frac{\pi}{16\sqrt{3}} \right ) L
- \frac{11915}{62208}
\right. \nonumber \\
&& \left.
   \quad + \frac{\pi}{24\sqrt{3}}S_2
         + \frac{\pi}{36 \sqrt{3}} + 
         \frac{39}{64}S_2  - \frac{1}{288}\zeta_3 + 
         \frac{347 \pi^2}{93312}
\right]
 \nonumber \\
&& 
       + \delta \left[  
  \left ( \frac{1}{72} - \frac{\pi}{72\sqrt{3}} \right )L 
- \frac{1}{1296} + \frac{5}{96}\frac{\pi}{\sqrt{3}}S_2 
       - \frac {11 \pi}{864 \sqrt{3}} 
\right. \nonumber \\
&& \left.
     \quad     - \frac{1}{384}S_2  - \frac{1}{216}\zeta_3 
 + \frac{53 \pi^2}{31104}  \right]
+ {\cal O}\left (\frac{m_\mu^4}{M^4},\delta^2 \right ),
\label{respi}
\ea
where $L = \log(M/m_\mu)$, $\zeta_3\simeq 1.202057$ is the Riemann zeta
function and $S_2 =
\frac{4}{9\sqrt{3}}\mathrm{Cl}_{2}\left(\frac{\pi}{3}\right) \simeq
0.260434$. 

Substituting $\alpha=1/137.036,~N_c=3$,
$M=769~{\rm MeV},~m_\mu=105.66~{\rm MeV},~m_\pi=
134.98~{\rm MeV}$, and 
$F_\pi=92.4~{\rm MeV}$ into Eq.~(\ref{respi0}) we obtain 
the result \cite{FullEq} in Eq.~(\ref{res}).

The result (\ref{res}) is free from numerical errors, which does not
mean that it represents the exact contribution of the pion pole to
$g_\mu-2$.  The dependence of that result on the model and form factor
parameters has been discussed in detail in \cite{Knecht:2001qf}.

\begin{figure}[htb]
\hspace*{-38mm}
\begin{minipage}{16.cm}
\begin{tabular}{c}
\psfig{figure=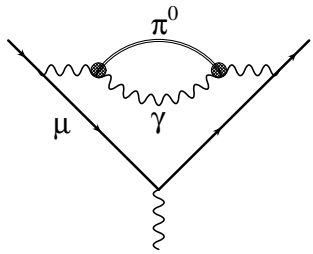,width=40mm}
\end{tabular}
\end{minipage}
\caption{Vacuum polarization contribution of the neutral pion.}
\label{fig:vacpol}
\end{figure}

Our result can be checked in several ways, in particular to ensure the
correct treatment of the Feynman rules in a computer code
\cite{form3com}.  For example, using the WZW Lagrangian one can
evaluate the vacuum polarization contribution to $g_\mu-2$ where the
virtual photon splits into $\pi^0$ and $\gamma$, shown in
Fig.~\ref{fig:vacpol}. The contribution of this diagram to $a_\mu$
should be positive, since it can be related via dispersion relations
to the cross section $\sigma(e^+e^-\to \pi^0\gamma)$
\cite{WMACunpub}. Indeed, we find
\be
a_\mu^{{\rm vp},\pi^0 \gamma}  = \left ( \frac{\alpha}{\pi} \right )^3 
\frac{m_\mu^2}{F_\pi^2} \left ( \frac{Nc}{\pi} \right )^2  X_{\pi^0 \gamma},
\ee
where 
\be
X_{\pi^0 \gamma} = 
 \frac{L}{1296}   + \frac{181}{15552}  - \frac{\pi}{96\sqrt{3}}
     + \frac{7 \pi^2}{7776}  
+ {\cal O}\left ( \frac{m_\mu^2}{M^2},\delta \right ).
\label{resvp}
\ee
Including several more terms in the $m_\mu/M$ and $\delta$ expansions, we
obtain $a_\mu^{{\rm vp},\pi^0 \gamma} \simeq 3.7 \times 10^{-11}$, a
positive contribution.

It is also useful to compute the pole contribution of a (hypothetical)
scalar meson ``$\sigma$'' with a mass close to $m_\mu$.  By analogy
with other diagrams which can be evaluated for scalar and pseudoscalar
contributions \cite{WM:Tennessee,MassMech,Czarnecki:2001pv}, one might
expect that the leading ultraviolet logarithm $L^2$ in the scalar
contribution has equal magnitude but opposite sign, compared to the
pseudoscalar $\pi^0$ \cite{WMpriv}.  We have confirmed this by an
explicit calculation, substituting $F^{\mu \nu}$ and $\sigma$ for
$\widetilde F^{\mu \nu}$ and $\pi^0$ in (\ref{wzw}), and we find
\ba
&& a_{\mu}^{{\rm LBL},\sigma} = \left ( \frac{\alpha}{\pi} \right )^3 
\frac{m_\mu^2}{F_\pi^2} \left ( \frac{Nc}{\pi} \right )^2  X_{\sigma},
%\ea
%\ba
\nonumber \\
\lefteqn{
X_{\sigma} = -{L^2\over 48}
+ {L\over 48} \,{\pi\over \sqrt{3}}
- {L\over 288}
+ {485 \over 10368} 
- {5 \over 96} \,{\pi\over \sqrt{3}} S_2 }
\nonumber \\
&&
- {29 \over 3456} \,{\pi\over \sqrt{3}} 
- {35 \over 15552} \pi^2 
+ {\zeta_3\over 288}
- {5 \over 128}  S_2
+ {\cal O}\left ( \frac{m_\mu^2}{M^2},\delta \right ).
\nonumber \\
\ea
We see that the leading logarithm is indeed the same as in
(\ref{respi}) except for the sign.

Let us also note that the calculation presented in this Letter can
easily be extended to provide analytic results for the pion pole
contribution when other models (see e.g. \cite{Knecht:2001qf}) for the
pion transition form factors are used.  Those other form factors can
be represented as linear combinations of heavy propagators $1/(q_1^2 -
M^2)$ and $1/(q_2^2-M^2)$, and the technique described in this Letter
is applicable.

Our result confirms the positive sign of the pion pole contribution to
the muon $g_\mu-2$.  We would like to note that this effect was found to
be positive already several years ago in a correct numerical
calculation \cite{kinoshita85}.  It was an apparently simple algebraic
mistake which led to the sign error in later works \cite{privTK}.  We
hope that our analytical result will help better understand the
dependence of this effect on the ultraviolet regulator ($M$) and thus
place the theory of the muon $g-2$ on firmer footing, as we await the
release of the new Brookhaven experimental result.

{\em Acknowledgments:} 
This research was supported in part by the Natural Sciences and
Engineering Research Council of Canada and by the DOE under grant
number DE-AC03-76SF00515.

%\bibliography{$HOME/pro/tex/phd}

\begin{thebibliography}{27}
\expandafter\ifx\csname natexlab\endcsname\relax\def\natexlab#1{#1}\fi
\expandafter\ifx\csname bibnamefont\endcsname\relax
  \def\bibnamefont#1{#1}\fi
\expandafter\ifx\csname bibfnamefont\endcsname\relax
  \def\bibfnamefont#1{#1}\fi
\expandafter\ifx\csname citenamefont\endcsname\relax
  \def\citenamefont#1{#1}\fi
\expandafter\ifx\csname url\endcsname\relax
  \def\url#1{\texttt{#1}}\fi
\expandafter\ifx\csname urlprefix\endcsname\relax\def\urlprefix{URL }\fi
\providecommand{\bibinfo}[2]{#2}
\providecommand{\eprint}[2][]{\url{#2}}

\bibitem[{\citenamefont{Brown et~al.}(2001)}]{Brown:2001mg}
\bibinfo{author}{\bibfnamefont{H.~N.} \bibnamefont{Brown}} \bibnamefont{et~al.}
  (\bibinfo{collaboration}{Muon g-2}), \bibinfo{journal}{Phys. Rev. Lett.}
  \textbf{\bibinfo{volume}{86}}, \bibinfo{pages}{2227} (\bibinfo{year}{2001}),
  \eprint[http://arXiv.org/abs]{hep-ex/0102017}.

\bibitem[{\citenamefont{Marciano and Roberts}(2001)}]{Marciano:2001qq}
\bibinfo{author}{\bibfnamefont{W.~J.} \bibnamefont{Marciano}} \bibnamefont{and}
  \bibinfo{author}{\bibfnamefont{B.~L.} \bibnamefont{Roberts}}
  (\bibinfo{year}{2001}), \eprint[http://arXiv.org/abs]{hep-ph/0105056}.

\bibitem[{\citenamefont{Melnikov}(2001)}]{Melnikov:2001uw}
\bibinfo{author}{\bibfnamefont{K.}~\bibnamefont{Melnikov}},
  \bibinfo{journal}{Int. J. Mod. Phys.} \textbf{\bibinfo{volume}{A16}},
  \bibinfo{pages}{4591} (\bibinfo{year}{2001}),
  \eprint[http://arXiv.org/abs]{hep-ph/0105267}.

\bibitem[{\citenamefont{Eidelman}(2001)}]{Eidelman:2001ju}
\bibinfo{author}{\bibfnamefont{S.~I.} \bibnamefont{Eidelman}},
  \bibinfo{journal}{Nucl. Phys. Proc. Suppl.} \textbf{\bibinfo{volume}{98}},
  \bibinfo{pages}{281} (\bibinfo{year}{2001}).

\bibitem[{\citenamefont{Jegerlehner}(2001)}]{Jegerlehner:2001wq}
\bibinfo{author}{\bibfnamefont{F.}~\bibnamefont{Jegerlehner}}
  (\bibinfo{year}{2001}), \eprint[http://arXiv.org/abs]{hep-ph/0104304}.

\bibitem[{\citenamefont{{H\"ocker}}(2001)}]{Hocker:2001fu}
\bibinfo{author}{\bibfnamefont{A.}~\bibnamefont{{H\"ocker}}}
  (\bibinfo{year}{2001}), \eprint[http://arXiv.org/abs]{hep-ph/0111243}.

\bibitem[{\citenamefont{Knecht and Nyffeler}(2001)}]{Knecht:2001qf}
\bibinfo{author}{\bibfnamefont{M.}~\bibnamefont{Knecht}} \bibnamefont{and}
  \bibinfo{author}{\bibfnamefont{A.}~\bibnamefont{Nyffeler}}
  (\bibinfo{year}{2001}), \eprint[http://arXiv.org/abs]{hep-ph/0111058}.

\bibitem[{\citenamefont{Knecht et~al.}(2001)\citenamefont{Knecht, Nyffeler,
  Perrottet, and {de Rafael}}}]{Knecht:2001qg}
\bibinfo{author}{\bibfnamefont{M.}~\bibnamefont{Knecht}},
  \bibinfo{author}{\bibfnamefont{A.}~\bibnamefont{Nyffeler}},
  \bibinfo{author}{\bibfnamefont{M.}~\bibnamefont{Perrottet}},
  \bibnamefont{and} \bibinfo{author}{\bibfnamefont{E.}~\bibnamefont{{de
  Rafael}}} (\bibinfo{year}{2001}),
  \eprint[http://arXiv.org/abs]{hep-ph/0111059}.

\bibitem[{\citenamefont{Hayakawa et~al.}(1995)\citenamefont{Hayakawa,
  Kinoshita, and Sanda}}]{Hayakawa:1995ps}
\bibinfo{author}{\bibfnamefont{M.}~\bibnamefont{Hayakawa}},
  \bibinfo{author}{\bibfnamefont{T.}~\bibnamefont{Kinoshita}},
  \bibnamefont{and} \bibinfo{author}{\bibfnamefont{A.~I.} \bibnamefont{Sanda}},
  \bibinfo{journal}{Phys. Rev. Lett.} \textbf{\bibinfo{volume}{75}},
  \bibinfo{pages}{790} (\bibinfo{year}{1995}), \eprint{hep-ph/9503463}.

\bibitem[{\citenamefont{Hayakawa et~al.}(1996)\citenamefont{Hayakawa,
  Kinoshita, and Sanda}}]{Hayakawa:1996ki}
\bibinfo{author}{\bibfnamefont{M.}~\bibnamefont{Hayakawa}},
  \bibinfo{author}{\bibfnamefont{T.}~\bibnamefont{Kinoshita}},
  \bibnamefont{and} \bibinfo{author}{\bibfnamefont{A.~I.} \bibnamefont{Sanda}},
  \bibinfo{journal}{Phys. Rev.} \textbf{\bibinfo{volume}{D54}},
  \bibinfo{pages}{3137} (\bibinfo{year}{1996}), \eprint{hep-ph/9601310}.

\bibitem[{\citenamefont{Hayakawa and Kinoshita}(1998)}]{Hayakawa:1998rq}
\bibinfo{author}{\bibfnamefont{M.}~\bibnamefont{Hayakawa}} \bibnamefont{and}
  \bibinfo{author}{\bibfnamefont{T.}~\bibnamefont{Kinoshita}},
  \bibinfo{journal}{Phys. Rev.} \textbf{\bibinfo{volume}{D57}},
  \bibinfo{pages}{465} (\bibinfo{year}{1998}), \eprint{hep-ph/9708227}.

\bibitem[{\citenamefont{Bijnens et~al.}(1995)\citenamefont{Bijnens, Pallante,
  and Prades}}]{Bijnens:1995cc}
\bibinfo{author}{\bibfnamefont{J.}~\bibnamefont{Bijnens}},
  \bibinfo{author}{\bibfnamefont{E.}~\bibnamefont{Pallante}}, \bibnamefont{and}
  \bibinfo{author}{\bibfnamefont{J.}~\bibnamefont{Prades}},
  \bibinfo{journal}{Phys. Rev. Lett.} \textbf{\bibinfo{volume}{75}},
  \bibinfo{pages}{1447} (\bibinfo{year}{1995}), \bibinfo{note}{erratum ibid.,
  {\bf 75}, 3781 (1995)}, \eprint{hep-ph/9505251}.

\bibitem[{\citenamefont{Bijnens et~al.}(1996)\citenamefont{Bijnens, Pallante,
  and Prades}}]{Bijnens:1996xf}
\bibinfo{author}{\bibfnamefont{J.}~\bibnamefont{Bijnens}},
  \bibinfo{author}{\bibfnamefont{E.}~\bibnamefont{Pallante}}, \bibnamefont{and}
  \bibinfo{author}{\bibfnamefont{J.}~\bibnamefont{Prades}},
  \bibinfo{journal}{Nucl. Phys.} \textbf{\bibinfo{volume}{B474}},
  \bibinfo{pages}{379} (\bibinfo{year}{1996}), \eprint{hep-ph/9511388}.

\bibitem[{\citenamefont{Chetyrkin}(1991)}]{Chetyrkin91}
\bibinfo{author}{\bibfnamefont{K.~G.} \bibnamefont{Chetyrkin}}
  (\bibinfo{year}{1991}), \bibinfo{note}{preprint MPI-Ph/PTh 13/91}.

\bibitem[{\citenamefont{Tkachev}(1994)}]{Tkachev:1994gz}
\bibinfo{author}{\bibfnamefont{F.~V.} \bibnamefont{Tkachev}},
  \bibinfo{journal}{Sov. J. Part. Nucl.} \textbf{\bibinfo{volume}{25}},
  \bibinfo{pages}{649} (\bibinfo{year}{1994}), \eprint{hep-ph/9701272}.

\bibitem[{\citenamefont{Smirnov}(1995)}]{Smirnov:1995tg}
\bibinfo{author}{\bibfnamefont{V.~A.} \bibnamefont{Smirnov}},
  \bibinfo{journal}{Mod. Phys. Lett.} \textbf{\bibinfo{volume}{A10}},
  \bibinfo{pages}{1485} (\bibinfo{year}{1995}), \eprint{hep-th/9412063}.

\bibitem[{\citenamefont{Fleischer et~al.}(1999)\citenamefont{Fleischer,
  Kalmykov, and Kotikov}}]{Fleischer:1999hp}
\bibinfo{author}{\bibfnamefont{J.}~\bibnamefont{Fleischer}},
  \bibinfo{author}{\bibfnamefont{M.~Y.} \bibnamefont{Kalmykov}},
  \bibnamefont{and} \bibinfo{author}{\bibfnamefont{A.~V.}
  \bibnamefont{Kotikov}}, \bibinfo{journal}{Phys. Lett.}
  \textbf{\bibinfo{volume}{B462}}, \bibinfo{pages}{169} (\bibinfo{year}{1999}),
  \eprint[http://arXiv.org/abs]{hep-ph/9905249}.

\bibitem[{Ful()}]{FullEq}
\bibinfo{note}{In Eq.~(\protect\ref{respi}) only a few terms in the expansion
  in $m/M$ and $\delta$ are presented; the complete result which we used for
  the numerical evaluation can be obtained from the authors.}

\bibitem[{for()}]{form3com}
\bibinfo{note}{For the computations in this paper we have employed FORM
  \cite{form3}. The Levi-Civita tensor in FORM is imaginary.}

\bibitem[{\citenamefont{Vermaseren}()}]{form3}
\bibinfo{author}{\bibfnamefont{J.~A.~M.} \bibnamefont{Vermaseren}},
  \bibinfo{note}{math-ph/0010025}.

\bibitem[{\citenamefont{Czarnecki and Marciano}(1999)}]{WMACunpub}
\bibinfo{author}{\bibfnamefont{A.}~\bibnamefont{Czarnecki}} \bibnamefont{and}
  \bibinfo{author}{\bibfnamefont{W.~J.} \bibnamefont{Marciano}}
  (\bibinfo{year}{1999}), \bibinfo{note}{unpublished}.

\bibitem[{\citenamefont{Marciano}(1995)}]{WM:Tennessee}
\bibinfo{author}{\bibfnamefont{W.~J.} \bibnamefont{Marciano}}, in
  \emph{\bibinfo{booktitle}{Radiative Corrections: Status and Outlook}}, edited
  by \bibinfo{editor}{\bibfnamefont{B.~F.~L.} \bibnamefont{Ward}}
  (\bibinfo{publisher}{World Scientific}, \bibinfo{address}{Singapore},
  \bibinfo{year}{1995}), pp. \bibinfo{pages}{403--414}.

\bibitem[{\citenamefont{Marciano}(1996)}]{MassMech}
\bibinfo{author}{\bibfnamefont{W.}~\bibnamefont{Marciano}}, in
  \emph{\bibinfo{booktitle}{Particle Theory and Phenomenology}}, edited by
  \bibinfo{editor}{\bibfnamefont{K.}~\bibnamefont{Lassila}}
  \bibnamefont{et~al.} (\bibinfo{publisher}{World Scientific},
  \bibinfo{address}{Singapore}, \bibinfo{year}{1996}), p.~\bibinfo{pages}{22}.

\bibitem[{\citenamefont{Czarnecki and Marciano}(2001)}]{Czarnecki:2001pv}
\bibinfo{author}{\bibfnamefont{A.}~\bibnamefont{Czarnecki}} \bibnamefont{and}
  \bibinfo{author}{\bibfnamefont{W.~J.} \bibnamefont{Marciano}},
  \bibinfo{journal}{Phys. Rev.} \textbf{\bibinfo{volume}{D64}},
  \bibinfo{pages}{013014} (\bibinfo{year}{2001}),
  \eprint[http://arXiv.org/abs]{hep-ph/0102122}.

\bibitem[{WMp()}]{WMpriv}
\bibinfo{note}{We are grateful to William Marciano for suggesting this test.}

\bibitem[{\citenamefont{Kinoshita et~al.}(1985)\citenamefont{Kinoshita, Nizic,
  and Okamoto}}]{kinoshita85}
\bibinfo{author}{\bibfnamefont{T.}~\bibnamefont{Kinoshita}},
  \bibinfo{author}{\bibfnamefont{B.}~\bibnamefont{Nizic}}, \bibnamefont{and}
  \bibinfo{author}{\bibfnamefont{Y.}~\bibnamefont{Okamoto}},
  \bibinfo{journal}{Phys. Rev.} \textbf{\bibinfo{volume}{D31}},
  \bibinfo{pages}{2108} (\bibinfo{year}{1985}).

\bibitem[{\citenamefont{Hayakawa and Kinoshita}()}]{privTK}
\bibinfo{author}{\bibfnamefont{M.}~\bibnamefont{Hayakawa}} \bibnamefont{and}
  \bibinfo{author}{\bibfnamefont{T.}~\bibnamefont{Kinoshita}},
  \bibinfo{note}{private communication}.

\end{thebibliography}

\end{document}